# Macroeconomic Forecasting using Dynamic Factor Models: The Case of Morocco, DAOUI, M.[*]

[*]Doctor of Economics and Management, Faculty of Law, Economics and Social Sciences, Interdisciplinary Research Laboratory in Economics, Finance and Management of Organizations, Sidi Mohamed Ben Abdellah University, Fez, Morocco, marouane.daoui@usmba.ac.ma.

**Abstrcat:**

This article discusses the use of dynamic factor models in macroeconomic forecasting, with a focus on the Factor-Augmented Error Correction Model (FECM). The FECM combines the advantages of cointegration and dynamic factor models, providing a flexible and reliable approach to macroeconomic forecasting, especially for non-stationary variables.

We evaluate the forecasting performance of the FECM model on a large dataset of 117 Moroccan economic series with quarterly frequency. Our study shows that FECM outperforms traditional econometric models in terms of forecasting accuracy and robustness.

The inclusion of long-term information and common factors in FECM enhances its ability to capture economic dynamics and leads to better forecasting performance than other competing models.

Our results suggest that FECM can be a valuable tool for macroeconomic forecasting in Morocco and other similar economies.

# Prévision macroéconomique à l'aide de modèles à facteurs dynamiques : Cas du Maroc

**Résumé :**

Cet article traite l'utilisation des modèles à facteurs dynamiques dans les prévisions macroéconomiques, en mettant l'accent sur le modèle à correction d'erreur augmenté de facteurs (FECM). Le FECM combine les avantages de la cointégration et des modèles à facteurs dynamiques, fournissant une approche flexible et fiable pour les prévisions macroéconomiques, en particulier pour les variables non stationnaires.

Nous évaluons la performance de prévision du modèle FECM sur un large ensemble de données de 117 séries économiques marocaines avec une fréquence trimestrielle.

Notre étude montre que le modèle FECM surpasse les modèles économétriques traditionnels en termes de précision et de robustesse des prévisions.

L'inclusion d'informations à long terme et de facteurs communs dans le modèle FECM améliore sa capacité à capturer la dynamique économique et conduit à de meilleures performances de prévision que d'autres modèles concurrents.

Nos résultats suggèrent que le FECM peut être un outil précieux pour les prévisions macroéconomiques au Maroc et dans d'autres économies similaires.

**Mots-clés:** Prévisions macroéconomiques, modèles à facteurs dynamiques, modèle à correction d'erreur augmenté de facteurs (FECM), cointégration, variables non stationnaires, Maroc.

**Introduction:**

Macroeconomic forecasting is an essential tool for policymakers, investors, and businesses to make informed decisions. However, traditional forecasting models have limitations, particularly in capturing the dynamic relationships between different economic variables. This is where dynamic factor models come in, which can provide a more accurate and reliable forecast by incorporating the dynamic interplay between multiple economic indicators.

Dynamic factor models are based on the idea that a relatively small number of unobserved factors that evolve over time are responsible for the common dynamics of a large number of time series variables. These models are increasingly used in economic forecasting because of their ability to capture the complex interrelationships among variables that affect the economy. Unlike traditional time series models, which rely solely on past observations to predict future outcomes, dynamic factor models use a large set of variables to capture the common factors that drive economic activity. By incorporating a wide range of information, including macroeconomic indicators and financial market data, dynamic factor models are able to provide more accurate forecasts of key economic variables.

Various researchers have extensively studied dynamic factor models, and there are many surveys on the topic, including those by Bai and Ng (2008b), Bai and Wang (2016), and Barigozzi (2018). The benefits of dynamic factor models and VAR models are combined in factor augmented VAR (FAVAR) models. A FAVAR model builds its factors from a wide range of economic data. The aggregate impacts of economic variables on the variables of interest are then captured by include these components in a VAR model. Bernanke, Boivin, and Eliasz (2005) were indeed some of the first researchers to implement Factor Augmented VAR (FAVAR) models in their empirical research. Recently, Daoui and Benyacoub (2021) used this methodology (FAVAR) to examine how monetary policy shocks affected Morocco's economic growth.

However, FAVAR models have been developed in the context of modeling stationary variables. While it is well known that most macroeconomic series must be considered as non-stationary and that they are stationary only after differentiation. In this context, it may therefore be useful to combine the advantages of dynamic factor models (the inclusion of many variables) and error correction models (the inclusion of long-term or cointegrating relationships). Indeed, Banerjee and Marcellino (2009) and Banerjee, Marcellino, and Masten (2010) further developed the FAVAR model into the Factor-Augmented Error Correction Model (FECM). The model uses a large dataset of economic variables to build factors that capture the aggregate impacts of these variables on the variables of interest. The long-term or cointegrating relationships between the variables are then included in the model as error correction terms, which help to account for any deviations from the long-run equilibrium relationship.

The objective of this article is to take advantage of the benefits of dynamic factor models, while integrating the cointegration relationships between variables at the level, which can thus constitute an interesting path to follow to improve short-term forecasting tools. We therefore propose an evaluation of the predictive capacity of factor augmented error correction models (FECM) using a large database of Moroccan economic data.

To achieve this objective, we will develop a comparative analysis of macroeconomic forecasting models, focusing specifically on the FECM as a powerful tool that provides a flexible and reliable approach to macroeconomic forecasting. We hypothesize that the FECM will outperform traditional econometric models in terms of forecasting accuracy and robustness when applied to a large Moroccan economic database. Moreover, we expect that the inclusion of long-term information and common factors in the FECM will enhance its ability to capture economic dynamics and lead to better forecasting performance than other models. Overall, our study aims to demonstrate the potential of the FECM as a valuable tool for economic forecasting in Morocco and other similar economies.

In this article, we discuss the use of dynamic factor models in macroeconomic forecasting, focusing on the case of Morocco. In the first section, we review the literature on FECM, which combines the advantages of cointegration and dynamic factor models, in order to improve our understanding of this econometric model. In the second section, we explain the methodology behind the FECM and how it allows for the inclusion of non-stationary variables as well as an extended set of information on factors extracted from a large dataset. In the third section, we present our data and empirical application, where we evaluate the forecasting performance of the FECM model on 117 Moroccan economic series with quarterly frequency. Finally, in the fourth section, we discuss the results of the forecasting comparison, providing detailed information on the mean squared errors (MSE) of the competing models for different horizons and variables analyzed.

## 1. Review of literature

Banerjee and Marcellino (2009) proposed the factor-augmented error correction model (FECM) as a synthesis of two significant advancements in econometric analysis: cointegration (e.g. Engle & Granger, 1987; Johansen 1995) and large dynamic factor models (e.g. Forni, Hallin, Lippi, & Reichlin, 2000; Stock & Watson, 2002a, 2002b). The FECM combines these two approaches to better understand economic dynamics.

There has been considerable discussion in the literature about the challenges of modeling large systems of equations in which the full cointegrating space may be difficult to determine (e.g. Clements and Hendry, 1995). On the other hand, large dynamic factor models and factor augmented VARs (FAVARs, e.g. Bernanke, Boivin, & Eliasz, 2005; Stock & Watson, 2005) often focus on differences between variables in order to achieve stationarity. The FECM combines these approaches by using factors extracted from large datasets in their raw form as a representation of the non-stationary common trends, which are then modeled together with selected economic variables of interest that can cointegrate with the factors. In this sense, the FECM includes both ECM and FAVAR models and may produce improved results, particularly when the conditions for consistent factor and parameter estimation are met and cointegration is relevant.

Banerjee, Marcellino, and Masten (2014) compared the forecasting accuracy of the FECM to that of the ECM and the FAVAR. The relative performance of these models can vary depending on the variables being analyzed and the characteristics of the underlying processes generating the data, such as the extent and strength of cointegration, the level of lagged dependence in the models, and the forecasting horizon. However, in general, the FECM tends to outperform both the ECM and the FAVAR.

Banerjee, Marcellino, and Masten (2014) examined the application of the FECM for structural analysis. Building on a dynamic factor model for non-stationary data (Bai, 2004), they derived the moving-average representation of the FECM and demonstrated how it can be used to identify structural shocks and their transmission mechanism, employing techniques similar to those used in the structural VAR literature.

The FECM model is related to a framework that has been used recently to test for cointegration in panels (e.g. Bai, Kao, & Ng, 2009; Gengenbach, Urbain, & Westerlund, 2008). While the approaches may seem similar at first glance, there are several key differences. First, in panel cointegration, the size of the dataset (the set of variables being tested for cointegration) is fixed and the units of the panel (i.e. $1, 2, \ldots, N$) provide repeated information on the cointegration vectors. In contrast, the dataset in the FECM model is, in principle, infinite in size and driven by a finite number of common trends. Second, as a result of the first difference, the role of the factors (whether integrated or stationary) is also different in the two approaches. In the panel cointegration framework, the factors capture cross-sectional dependence but are not cointegrated with the vector of variables of interest. In the FECM model, cointegration between the variables and the factors is allowed and serves as a proxy for the missing cointegrating information in the entire dataset.

Another related but distinct paper is Barigozzi, Lippi, and Luciani (2014). They use a non-parametric static version of the factor model with common $I(1)$ factors only, while in the FECM model, the factors can be both $I(1)$ and $I(0)$, which adds complexity, particularly for structural applications involving permanent shocks (see Banerjee et al., 2014b). They also assume that the factors follow a VAR model and demonstrate that their first differences can be represented using a finite-order ECM model, which is a noteworthy finding. In contrast, the focus in the FECM model is on cointegration between the factors and the observable variables. The Barigozzi et al. (2014) model is also similar to the one analyzed by Bai (2004). Bai did not consider impulse responses, but it would be possible to obtain consistent estimates of the responses based on the factors in their raw form (rather than using the ECM model for the factors as in Barigozzi et al., 2014) within the context of his model.

Recently, Benyacoub and Daoui (2021) applied the FECM to investigate the impact of monetary policy shocks on economic growth in Morocco. They found that the model produces improved results, particularly when the conditions for consistent factor and parameter estimation are met and cointegration is relevant.

In this article, we demonstrate how the Factor Augmented Error Correction Model (FECM) can be used for forecasting. We provide a novel empirical case utilizing a large Moroccan economic database to demonstrate the application of FECM.

## 2. Methodology

To model a vector of non-stationary series in VAR form, and when these series are cointegrated, Johansen (1995) shows that the correct specification for this modeling is a representation called a vector error correction model (VECM) that describes the links between the contemporaneous values of the differentiated series, their lagged values, and the deviations from the long-run relationships (the cointegrating relationships) measured in the previous period.

Thus, when a vector $y_t$ is non-stationary but cointegrated, with a cointegrating matrix $\beta$, and if $y_t$ admits a VAR representation of order $p$ of the following form:

$$y_t = \mu + \sum_{k=1}^{p} A_k y_{t-k} + \epsilon_t \qquad (1)$$

We show that it also admits a VECM representation of the following form :

$$\Delta y_t = \mu - \alpha \beta' y_{t-1} + \sum_{k=1}^{p-1} \Phi_k \Delta y_{t-k} + \epsilon_t \qquad (2)$$

From this VECM representation (equation 2), we can see that it involves only stationary variables (the differentiated series and the cointegration relationships) while retaining, via the cointegration relationships, information on the levels of the variables. Conversely, the VAR representation (equation 1), which would only consider the dynamic relationships between the differentiated variables, would be poorly specified and would not consider all the available information.

Banerjee and Marcellino (2009) and Banerjee, Marcellino, and Masten (2010) have extended the FAVAR model to the Factor augmented Error Correction Model (FECM), which allows for the inclusion of non-stationary variables. These models allow, as in the FAVAR models, the inclusion of an extended set of information via factors extracted from a large data set. They also allow, as in the VECM models, to consider information on the levels of the variables, via cointegration relationships.

Thus, a possible specification of the Factor augmented Error Correction Model (FECM) is as follows:

$$\begin{bmatrix} \Delta y_t \\ \Delta f_t \end{bmatrix} = \begin{bmatrix} \alpha_y \\ \alpha_f \end{bmatrix} \beta' \begin{bmatrix} y_{t-1} \\ f_{t-1} \end{bmatrix} + \Phi_1 \begin{bmatrix} \Delta y_{t-1} \\ \Delta f_{t-1} \end{bmatrix} + \cdots + \Phi_p \begin{bmatrix} \Delta y_{t-p} \\ \Delta f_{t-p} \end{bmatrix} + \begin{bmatrix} \epsilon_{y_t} \\ \epsilon_{f_t} \end{bmatrix} \qquad (3)$$

where $y_t$ is the vector of small-dimensional variables of interest and $f_t$ is the vector of factors extracted from a large data set also of small dimension.

We also present the most widely used methodology in the empirical literature on FECM models, namely the methodology of Banerjee et al. (2008, 2010) which has several practical steps.

Starting with the extraction of the non-stationary factors, Banerjee et al. (2008, 2010) proposes two approaches. The first one consists in extracting the $I(1)$ factors using only the $I(1)$ series of the base constituted by all the non-stationary variables. In the second approach, the cumulative values of the stationary series are used.[1] However, they show that whatever method is used, the results are very similar.

Subsequently, to specify the number of factors, these same authors suggested using the criteria proposed by Bai (2004) to determine the number of $I(1)$ factors and the criteria of Bai and Ng (2002) to select the number of $I(0)$ factors. The $I(1)$ factors are estimated using the method advocated by Bai (2004), i.e. using PCA on the level series.[2] However, the stationary factors are estimated by applying a PCA to the full base, in which the non-stationary series are replaced by their first differences.

However, since the number of $I(1)$ factors and the number of $I(0)$ factors obtained are in practice not compatible in the simulation results of Banerjee et al. (2008, 2010). They suggest that the level $I(1)$ variables may have both common trends and common cycles: this would be a case where the $I(1)$ variables have both common $I(1)$ factors and common $I(0)$ factors, which is quite plausible. In this case, Banerjee et al. (2008, 2010) first estimate the $I(1)$ factors and then compute the residuals of the regression of the $I(1)$ variables on the $I(1)$ factors. They then conduct a PCA on these residuals, which are assumed to be stationary, to extract new common stationary factor(s) that can eventually be added to the previous stationary factors.

Finally, the stationary and non-stationary factors are included in the FECM model, whose number of lags is determined by minimizing the Hannan-Quin information criterion. This model is estimated as a usual VECM model and, under the assumptions of Bai (2004), replacing the unobservable factors by the estimated factors does not change the properties of the obtained estimators. The model is then used to make forecasts of the variables of interest $y_t$.

## 3. Data and empirical application

In this section, we illustrate the performance of the FECM in forecasting. In order to provide convincing evidence about the usefulness of the FECM model as a forecasting tool, we evaluate the forecasting performance of the FECM model, which includes as variables the Producer Price Index (**PPI**), the Consumer Price Index (**CPI**), and the Money Market Interest Rate (**MMIR**). In fact, our database contains 117 Moroccan economic series with a quarterly frequency (from 1985:Q1 to 2018:Q4). The variables used in this study were mainly obtained from statistical data and reports provided by sources such as Bank Al-Maghrib, the High Commissioner for Planning, the Exchange Office, the Casablanca Stock Exchange and the International Monetary Fund. The full set of economic series used and their descriptions are presented in Table 1. In order to make them stationary, the variables are transformed on the basis of the transformation codes (TC) listed below:

(1) - No transformation: $X_{it} = Y_{it}$

(2) - First difference: $X_{it} = \Delta Y_{it}$

(3) - Second difference: $X_{it} = \Delta^2 Y_{it}$

(4) – Logarithm: $X_{it} = \log Y_{it}$

(5) - First difference of logarithm: $X_{it} = \Delta \log Y_{it}$

(6) - Second difference of logarithm: $X_{it} = \Delta^2 \log Y_{it}$

---

[1] The cumulative values of a series $x_{it}$ are defined by $X_{it} = \sum_{s=1}^{t} x_{is}$.
[2] It should be noted that the authors do not specify whether they controlled for the stationarity of the idiosyncratic components, which is a sine qua non condition for the validity of the results of Bai (2004).

**Table 1: Data Description**

| # | Mnemonic | Description | TC |
|---|---|---|---|
| **Gross Domestic Product by branch of activity, base 2007 (MMAD)** | | | |
| 1 | TGDP | Gross Domestic Product (Total) | 5 |
| 2 | GDPFIA | GDP: Financial and insurance activities | 5 |
| 3 | GDPPA | GDP: Primary Activities | 5 |
| 4 | GDPGPASS | GDP: General public administration and social security | 5 |
| 5 | GDPSA | GDP: Secondary activities | 5 |
| 6 | GDPONFS | GDP: Other Non-Financial Services | 5 |
| 7 | GDPTA | GDP: Tertiary activities | 5 |
| 8 | GDPBPW | GDP: Building and Public Works | 5 |
| 9 | GDPT | GDP: Trade | 5 |
| 10 | GDPEGW | GDP: Electricity, gas and water | 5 |
| 11 | GDPEHSA | GDP: Education, Health and Social Action | 5 |
| 12 | GDPEP | GDP : excluding primary | 5 |
| 13 | GDPHR | GDP: Hotels and restaurants | 5 |
| 14 | GDPEI | GDP: Extraction industry | 5 |
| 15 | GDPRERBS | GDP: Real estate, rentals and business services | 5 |
| 16 | GDPMI | GDP: Manufacturing industry | 5 |
| 17 | GDPTPS | GDP: Taxes on products net of subsidies | 5 |
| 18 | GDPPT | GDP: Post and telecommunications | 5 |
| 19 | GDPTI | GDP: Total Industries | 5 |
| 20 | GDPT | GDP: Transportation | 5 |
| 21 | GDPNA | GDP: non-agricultural AV | 5 |

| | | National production by branch of activity, base 2007 (MMAD) | |
|---|---|---|---|
| 22 | NPTBL | NP: Total business lines | 5 |
| 23 | NPGPASS | NP: General Public Administration and Social Security | 5 |
| 24 | NPFIA | NP: Financial and Insurance Activities | 5 |
| 25 | NPPA | NP: Primary Activities | 5 |
| 26 | NPONFS | NP: Other Non-Financial Services | 5 |
| 27 | NPBPW | NP: Building and Public Works | 5 |
| 28 | NPT | NP: Trade | 5 |
| 29 | NPEGW | NP: Electricity, gas and water | 5 |
| 30 | NPEHSA | NP: Education, Health and Social Action | 6 |
| 31 | NPHR | NP: Hotels and Restaurants | 5 |
| 32 | NPEI | NP: Extraction Industry | 5 |
| 33 | NPRERBS | NP: Real Estate, Rental and Business Services | 6 |
| 34 | NPMI | NP: Manufacturing Industry | 5 |
| 35 | NPPT | NP: Post and Telecommunications | 5 |
| 36 | NPRPOEP | NP: Refining of Petroleum and Other Energy Products | 2 |
| 37 | NPT | NP: Transportation | 5 |
| | | **Gross National Disposable Income, base 2007 (In million MAD)** | |
| 38 | FCE | Final consumption expenditure | 5 |
| 39 | FCEGG | Final consumption expenditure general government | 5 |
| 40 | HFCE | Household final consumption expenditure | 2 |
| 41 | GNI | Gross National Income | 2 |
| 42 | GNDI | Gross National Disposable Income | 2 |
| 43 | NPIES | Net property income from external sources | 2 |
| 44 | NCTA | Net current transfers from abroad | 5 |
| | | **Gross National Expenditure** | |
| 45 | GNED | Gross national expenditure deflator | 2 |
| 46 | GNECU | Gross national expenditure (units of local currency in current) | 5 |
| 47 | GNECO | Gross national expenditure (local currency units in constant dollars) | 5 |
| 48 | GNEGDP | Gross national expenditure (% of GDP) | 2 |
| | | **Final consumption expenditure** | |
| 49 | FCECU | Final consumption expenditure (current MAD) | 5 |
| 50 | FCECO | Final consumption expenditure (constant MAD) | 5 |
| 51 | HDCCU | Household final consumption at current prices | 5 |
| 52 | FCEHNGDP | Final consumption expenditure of households and NPISHs (% of GDP) | 2 |
| 53 | FCENNCU | Final consumption expenditure, nominal, national currency | 5 |
| 54 | FCENGDP | Final consumption expenditure, nominal, ratio to GDP, percentage | 2 |
| 55 | FCEPSN | Final consumption expenditure, private sector, nominal, MAD | 2 |
| 56 | FCEGDP | Final consumption expenditure (% of GDP) | 2 |
| 57 | FCEPS | Final consumption expenditure, public sector, MAD | 5 |

| | | **Investments** | |
|---|---|---|---|
| 58 | INVR | Investment rate | 2 |
| 59 | GIR | Gross investment rate | 2 |
| 60 | FDINI | Foreign direct investment, net inflows (% of GDP) | 2 |
| 61 | FDINO | Foreign direct investment, net outflows (% of GDP) | 2 |
| 62 | GFCFGR | Gross fixed capital formation (% growth) | 2 |
| 63 | GFCFCO | Gross fixed capital formation (constant MAD) | 5 |
| 64 | GFCFCU | Gross fixed capital formation (current MAD) | 5 |
| | | **Currency (In MMAD)** | |
| 65 | M1 | M1 | 5 |
| 66 | M2 | M2 | 5 |
| 67 | M3 | M3 | 5 |
| 68 | ORA | Official reserve assets | 2 |
| 69 | BCC | Banknotes and coins in circulation | 5 |
| 70 | TACVB | Term accounts and cash vouchers with banks | 2 |
| 71 | SAB | Savings accounts with banks | 4 |
| 72 | FCIR | Fiduciary Circulation | 5 |
| 73 | NRBAM | Net receivables from BAM | 2 |
| 74 | REC | Receivables | 2 |
| 75 | DDEP | Demand deposit | 2 |
| 76 | SDBAM | Sight deposits with BAM | 2 |
| 77 | SDBKS | Sight deposits with banks | 2 |
| 78 | DDEPTR | Demand deposits with the Treasury | 2 |
| 79 | CBKS | Cash at banks | 2 |
| 80 | COMM | Commitments | 2 |
| 81 | SINV | Sight Investments | 5 |
| 82 | NIRES | Net International Reserves | 2 |
| 83 | OMASS | Other monetary assets | 2 |
| | | **Stock market indicators** | |
| 84 | SETUR | Stock exchange turnover (In MMAD) | 2 |
| 85 | MCAP | Market capitalization (in millions of MAD) | 2 |
| 86 | DIVD | Dividends (In Millions of MAD) | 5 |
| | | **Gross national savings, 2007 base (in millions of Moroccan dirhams)** | |
| 87 | GNS | Gross national savings | 2 |
| 88 | GDS | Gross domestic savings | 2 |
| 89 | EXTS | External savings | 2 |
| | | **Inflation, Consumer Price Index** | |
| 90 | INFGDP | Inflation, GDP deflator (in %) | 2 |
| 91 | INFCP | Inflation, consumer prices (in %) | 1 |
| 92 | CPI | Consumer Price Index (2010 = 100) | 4 |
| 93 | CPIPY | CPI, Corresponding period of the previous year (in %) | 2 |
| 94 | CPOPP | CPI, Previous period (in %) | 1 |
| **Industrial producer price index** | | | |
| 95 | IPPIMAN | Industrial producer price index, manufacturing | 2 |
| 96 | IPPIMIN | Industrial producer price index, mining | 2 |
| **Unemployment rate** | | | |
| 97 | UMPR | Unemployment rate | 2 |
| **Exchange rates** | | | |
| 98 | ILTREGFE* | International Liquidity, Total reserves excluding gold, foreign exchange, SDR (Million) | 2 |
| 99 | ILTREGFC* | International Liquidity, Total reserves excluding gold, foreign currencies, US dollars. (Million) | 2 |
| 100 | ERMADSDR* | Exchange rate, MAD per SDR, average for the period | 2 |
| 101 | ERMADEUR* | Exchange rates, MAD per euro, Average for the period | 1 |
| 102 | ERMADUSD* | Exchange rate, MAD per US Dollar, average for the period | 2 |
| 103 | NEERI* | Exchange rate, Nominal effective exchange rate, Index | 2 |
| **Interest rates** | | | |
| 104 | IRDEP* | Interest rate, Deposit, | 2 |
| 105 | IRDIS* | Interest rate, discount | 2 |
| 106 | MMIR* | Interest rate, money market | 2 |
| 107 | IRGBYSMT* | Interest rates, government bond yields, short and medium term % %. | 2 |
| **Foreign trade** | | | |
| 108 | EBGS | External balance of goods and services (% of GDP) | 2 |
| 109 | TRAGDP | Trade (% of GDP) | 2 |
| 110 | EGSCU | Exports of goods and services (current MAD) | 2 |
| 111 | EGSCO | Exports of goods and services (constant MAD) | 2 |
| 112 | EGSGDP | Exports of goods and services (% of GDP) | 2 |
| 113 | IMPGSCU | Imports of goods and services (current MAD) | 2 |
| 114 | IMPGSCO | Imports of goods and services (constant MAD) | 2 |
| 115 | IMPGSGDP | Imports of goods and services (% of GDP) | 2 |
| 116 | EXTBGSCU | External balance in goods and services (current MAD) | 2 |
| 117 | EXTBGSCO | External balance in goods and services (constant MAD) | 2 |

**- Source:** Author.

In order to estimate the $I(1)$ factors to be utilized in the FECM, we selected 114 series that were deemed to be $I(1)$ and cumulated them with the remaining three $I(0)$ series before extracting the $I(1)$ factors. Prior to computing the factors and estimating the competing forecasting models, the raw data underwent several transformations.

Specifically, we applied a natural logarithm to all time series except interest rates. To obtain stationarity for the extraction of the $I(0)$ factors used in the FAVAR analysis, all series were differentiated once. If not already adjusted at source, time series were checked for the presence of seasonal components and adjusted accordingly with the $X-11$ filter before the forecast simulations.

Large outliers are observations that deviate from the sample median by more than six times the sample interquartile range. Following Stock and Watson (2005), the identified outliers were set to the median of the previous five observations.

To compute the $I(1)$ factors included in the FECM, all variables are treated as $I(1)$ with a non-zero mean and are estimated using the method of Bai (2004). For the $I(0)$ factors included in the FAVAR and FAR (factor augmented autoregressive model) models, we first transform the data into stationary series and then use the principal components-based estimator of Stock and Watson (2002a).

Forecasts are made using several concurrent models. First, we use three models, all based on observable variables only: an autoregressive (AR) model, a vector autoregressive (VAR) model, and an error correction model (ECM). To assess the role of additional information in forecasting, the second set of models complements the first set with factors extracted from the larger set of available variables: the FAR, FAVAR, and FECM specifications are AR, VAR, and ECM models augmented with factors, respectively.

For the FECM, we use two approaches for factor extraction. Our first choice is PCA estimation from the level data. As a robustness check, we use the factors estimated by the method of Bai and Ng (2004) applied to differentiated data: such an FECM is called FECMc.

The number of factors $I(1)$ and $I(0)$ is 1 and 3 respectively, and they are kept fixed over the forecast period, but their estimates are updated recursively. The lag lengths are determined by the BIC information criterion.[3] As for the cointegration test to determine the cointegrating ranks of the ECM and the FECM, we considered two approaches: the Johansen (1995) trace test and the Cheng and Phillips (2009) semiparametric test based on the BIC information criterion. Both methods yielded very similar results, but we preferred the Cheng and Phillips method due to its lower computational complexity and ease of implementation in practice.[4]

All variables are treated at the $I(1)$ level with a deterministic trend, which means that the dynamic forecasts of the differences of the variables (in logs) at a horizon $h$ produced by each of the competing models are aggregated to obtain the forecasts at the h-horizon level. We consider four different forecast horizons $h = 1, 2, 4, 8$. In contrast to our use of iterated forecasts at horizons $h$ (dynamic forecasts), Stock and Watson (1988) and Stock and Watson (2002a) use direct forecasts at horizon $h$, while Marcellino, Stock, and Watson (2006) find that iterated forecasts are often better except in the presence of misspecification.

---

[3] We also used the Hannan-Quinn (HQ) criterion to verify and confirm the robustness of the results.
[4] According to the simulation results of Cheng and Phillips (2009), using the BIC information criterion tends to underestimate the rank when the actual rank is low, while it performs best when the actual cointegrating rank is very low (0 or 1). Since model selection by the BIC information criterion is generally preferred for forecasting model selection, we chose to use it to test the cointegration rank as well. However, our results are also robust to the use of the Hannan-Quinn (HQ) criterion.

In the FECM framework, these forecasts are easier to make than their $h$-horizon counterparts. Similarly, the direct and iterative forecasting methods produce similar benchmark results at $h$ horizons on a common estimation and evaluation sample.

## 4. Results and discussion

The results of the forecast comparison are presented in Figure 1 and Tables 2, 3, 4 and 5. These tables provide detailed information on the MSEs of the competing models compared to the MSEs of the AR model for different horizons (1, 2, 4, and 8) and for each variable analyzed (PPI, CPI, and MMIR). The tables also provide information on the choice of cointegration rank and the number of lags in each model.

**Figure 1: RMSE comparison of the different models versus the RMSE of the AR model for the evaluation period [2012:Q1 - 2018:Q4]**

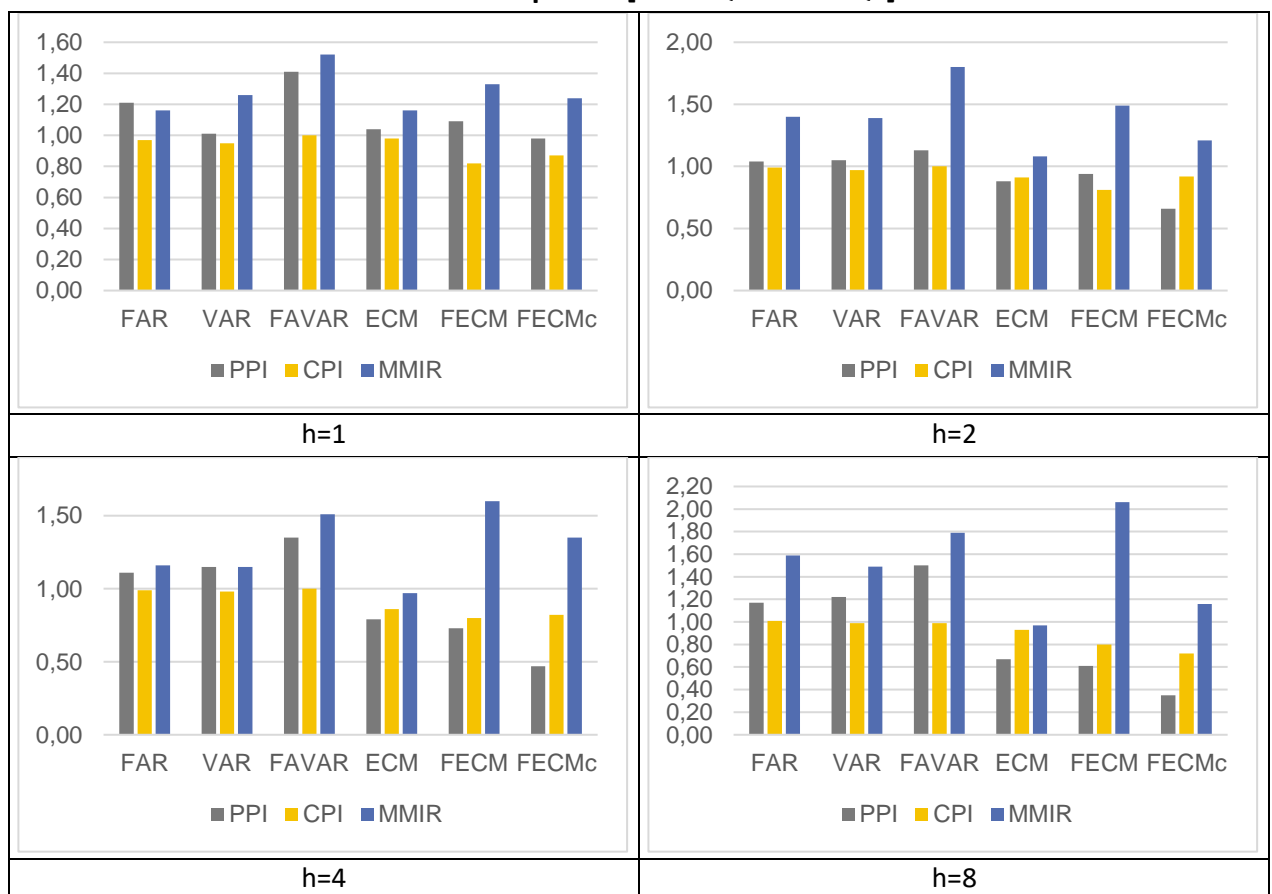

**Source:** Author's simulation results.

Looking at Tables 2, 3, 4, and 5, we can see that the FECM (and FECMc) outperform all other models for predicting the PPI and CPI variables at all horizons up to h = 8. In fact, the FECM (or FECMc) is the best performing model in all eight cases presented. The ECM, while not as good as the FECM, still performs significantly better than the FAR, VAR, and FAVAR for the MMIR variable. These results highlight the importance of the two factors and cointegration in forecasting this system.

**Table 2: RMSE of the different models compared to the RMSE of the AR model for the evaluation period [2012: Q1 - 2018: Q4] for the horizon h=1**

| Variables | RMSE of AR | MSE model/MSE_AR | | | | | |
|---|---|---|---|---|---|---|---|
| | | FAR | VAR | FAVAR | ECM | FECM | FECMc |
| PPI | 0,93 | 1,21 | 1,01 | 1,41 | 1,04 | 1,09 | **0,98** |
| CPI | 0,75 | 0,97 | 0,95 | 1,00 | 0,98 | **0,82** | 0,87 |
| MMIR | 1,26 | **1,16** | 1,26 | 1,52 | **1,16** | 1,33 | 1,24 |
| **Lags** | AR | 3,00 | 0,00 | 4,00 | | | |
| | FAR | 1,14 | 1,97 | 1,55 | | | |
| | VAR | FAVAR | ECM | FECM | | | |
| | 1,00 | 0,00 | 0,69 | 0,00 | | | |
| **Cointegration rank** | | ECM | | | FECM | | |
| | | Mean | Min | Max | Mean | Min | Max |
| | | 2,00 | 2,00 | 2,00 | 2,55 | 2,00 | 3,00 |

- **Source:** Author's calculations.

- **Notes:** Cheng and Phillips (2009) cointegration test and choice of lag number are based on the BIC criterion.
- **Data:** 1985: Q1 - 2018: Q4.
- **Forecasts:** 2012: Q1 - 2018: Q4.
- **Variables:** Producer Price Index (PPI), Consumer Price Index (CPI) and Money Market Interest Rate (MMIR).

**Table 3: RMSE of the different models compared to the RMSE of the AR model for the evaluation period [2012: Q1 - 2018: Q4] for the horizon h=2**

| Variables | RMSE of AR | MSE model/MSE_AR | | | | | |
|---|---|---|---|---|---|---|---|
| | | FAR | VAR | FAVAR | ECM | FECM | FECMc |
| PPI | 0,85 | 1,04 | 1,05 | 1,13 | 0,88 | 0,94 | **0,66** |
| CPI | 0,69 | 0,99 | 0,97 | 1,00 | 0,91 | **0,81** | 0,92 |
| MMIR | 1,55 | 1,40 | 1,39 | 1,80 | **1,08** | 1,49 | 1,21 |
| **Lags** | AR | 3,00 | 0,00 | 4,00 | | | |
| | FAR | 1,07 | 0,97 | 1,48 | | | |
| | VAR | FAVAR | ECM | FECM | | | |
| | 1,00 | 0,00 | 0,66 | 0,00 | | | |
| **Cointegration rank** | | ECM | | | FECM | | |
| | | Mean | Min | Max | Mean | Min | Max |
| | | 2,00 | 2,00 | 2,00 | 2,55 | 2,00 | 3,00 |

- **Source:** Author's calculations.

- **Notes:** Cheng and Phillips (2009) cointegration test and choice of lag number are based on the BIC criterion.
- **Data:** 1985: Q1 - 2018: Q4.
- **Forecasts:** 2012: Q1 - 2018: Q4.
- **Variables:** Producer Price Index (PPI), Consumer Price Index (CPI) and Money Market Interest Rate (MMIR).

**Table 4: RMSE of the different models compared to the RMSE of the AR model for the evaluation period [2012: Q1 - 2018: Q4] for the horizon h=4**

| Variables | RMSE of AR | MSE model/MSE_AR | | | | | |
|---|---|---|---|---|---|---|---|
| | | FAR | VAR | FAVAR | ECM | FECM | FECMc |
| PPI | 1,05 | 1,11 | 1,15 | 1,35 | 0,79 | 0,73 | **0,47** |
| CPI | 0,24 | 0,99 | 0,98 | 1,00 | 0,86 | **0,80** | 0,82 |
| MMIR | 1,96 | 1,16 | 1,15 | 1,51 | **0,97** | 1,60 | 1,35 |
| **Lags** | AR | 3,00 | 0,00 | 4,00 | | | |
| | FAR | 1,00 | 0,97 | 1,34 | | | |
| | VAR | FAVAR | ECM | FECM | | | |
| | 1,00 | 0,00 | 0,59 | 0,00 | | | |
| **Cointegration rank** | | ECM | | | FECM | | |
| | | Mean | Min | Max | Mean | Min | Max |
| | | 2,00 | 2,00 | 2,00 | 2,55 | 2,00 | 3,00 |

- **Source:** Author's calculations.

- **Notes:** Cheng and Phillips (2009) cointegration test and choice of lag number are based on the BIC criterion.
- **Data:** 1985: Q1 - 2018: Q4.
- **Forecasts:** 2012: Q1 - 2018: Q4.
- **Variables:** Producer Price Index (PPI), Consumer Price Index (CPI) and Money Market Interest Rate (MMIR).

**Table 5: RMSE of the different models compared to the RMSE of the AR model for the evaluation period [2012: Q1 - 2018: Q4] for the horizon h=8**

| Variables | RMSE of AR | MSE model/MSE_AR | | | | | |
|---|---|---|---|---|---|---|---|
| | | FAR | VAR | FAVAR | ECM | FECM | FECMc |
| PPI | 1,43 | 1,17 | 1,22 | 1,50 | 0,67 | 0,61 | **0,35** |
| CPI | 0,15 | 1,01 | 0,99 | 0,99 | 0,93 | 0,80 | **0,72** |
| MMIR | 2,81 | 1,59 | 1,49 | 1,79 | **0,97** | 2,06 | 1,16 |
| **Lags** | AR | 3,00 | 0,00 | 4,00 | | | |
| | FAR | 1,00 | 1,00 | 1,07 | | | |
| | VAR | FAVAR | ECM | FECM | | | |
| | 1,00 | 0,03 | 0,45 | 0,00 | | | |
| **Cointegration rank** | | ECM | | | FECM | | |
| | | Mean | Min | Max | Mean | Min | Max |
| | | 2,00 | 2,00 | 2,00 | 2,62 | 2,00 | 3,00 |

- **Source:** Author's calculations.

- **Notes:** Cheng and Phillips (2009) cointegration test and choice of lag number are based on the BIC criterion.
- **Data:** 1985: Q1 - 2018: Q4.
- **Forecasts:** 2012: Q1 - 2018: Q4.
- **Variables:** Producer Price Index (PPI), Consumer Price Index (CPI) and Money Market Interest Rate (MMIR).

However, for the financial variable (MMIR), FECM and FECMc do not provide the best results, with ECM performing best in three out of four cases (and FAR and ECM tied for h = 1). This suggests that the use of long-term information for forecasting the financial variable (MMIR) is limited.

These results are consistent with the findings of Clements and Hendry (1998) that forecast biases due to the omission of error correction terms are more significant at shorter horizons. Thus, the larger gains in FECM forecast accuracy at shorter horizons are consistent with their study.

Overall, the results suggest that the FECM (and FECMc) are the most effective models for forecasting the PPI and CPI variables, while the ECM is the most effective for the MMIR variable. Understanding the relative performance of these models can help inform future forecasting efforts in this system.

**Conclusion and prospects:**

In conclusion, this article highlights the benefits of Dynamic Factor Models, particularly the Factor-Augmented Error Correction Model, as a powerful tool for macroeconomic forecasting. The FECM model combines the advantages of cointegration and dynamic factor models, providing a flexible and reliable approach to macroeconomic forecasting, especially in the case of non-stationary variables.

The study showed that FECM outperforms traditional econometric models in terms of forecasting accuracy and robustness when applied to a large Moroccan economic database. The inclusion of long-term information and common factors in FECM enhanced its ability to capture economic dynamics and lead to better forecasting performance than other models. Specifically, the results showed that the FECM outperformed all other models in predicting selected economic variables such as the PPI, CPI, and MMIR at all study horizons.

However, the study also had some limitations. Firstly, the focus was on the Moroccan economy, which may limit the generalizability of the findings to other countries. Secondly, the study only used quarterly data, which may not capture short-term fluctuations in the economy. Future research could address these limitations by using data from other countries and incorporating higher frequency data.

Overall, the study contributes to the literature on dynamic factor models and their applications in macroeconomic forecasting. The results suggest that FECM is a powerful tool that provides a flexible and reliable approach to macroeconomic forecasting. It can be used by policymakers to make informed decisions and improve their understanding of the complex interrelationships among economic variables.

Prospects for future research include the use of machine learning techniques and artificial intelligence can be used to improve the performance of the model and to make it more robust. Furthermore, comparing the performance of the model in different business cycles and recessionary periods can provide valuable insights into the model's robustness.